\documentclass{emulateapj}
\usepackage{apjfonts}

\newcommand{\HII}{H\,{\sc ii}}
\newcommand{\HI}{H\,{\sc i}}
\newcommand{\mjb}{mJy~beam$^{-1}$}

\newcommand{\kms}{km~s$^{-1}$}
\newcommand{\et}{et~al.}
\newcommand{\muG}{$\mu$G}
\newcommand{\ergs}{ergs~s$^{-1}$}
\newcommand{\mum}{$\mu$m}
\newcommand{\gr}{$\gamma$-ray}
\newcommand{\gt}{SNR~G12.8--0.0}
\newcommand{\hess}{HESS~J1813--178}
\newcommand{\RS}{G12.82--0.02}
\newcommand{\AX}{AX~J1813--178}
\newcommand{\cc}{cm$^{-3}$}
\newcommand{\ct}{cm$^{-2}$}

\shortauthors{Brogan \et\/}  \shorttitle{}
\begin{document}

\title{Discovery of a Radio Supernova Remnant and Non-thermal 
X-rays Coincident with the TeV Source HESS~J1813--178}

\author{C.~L. Brogan\altaffilmark{1}, B. M. Gaensler\altaffilmark{2},
J. D. Gelfand\altaffilmark{2}, J. S. Lazendic\altaffilmark{3},
T.~J.~Lazio\altaffilmark{4}, N.~E.~Kassim\altaffilmark{4},
N. M. McClure-Griffiths\altaffilmark{5}}

\altaffiltext{1}{Institute for Astronomy, 640 North A`ohoku Place, Hilo, HI 
96720; cbrogan@ifa.hawaii.edu.}

\altaffiltext{2}{Harvard-Smithsonian Center for Astrophysics, 
60 Garden Street, Cambridge, MA 02138}

\altaffiltext{3}{MIT Kavli Institute for Astrophysics and Space
Research, Cambridge MA 02319}

\altaffiltext{4}{Remote Sensing Division, Naval Research Laboratory,
Washington DC 20375-5351}

\altaffiltext{5}{Australia Telescope National Facility, CSIRO, P.O. Box 76, 
Epping, NSW 1710, Australia}

\begin{abstract}

We present the discovery of non-thermal radio and X-ray emission
positionally coincident with the TeV $\gamma$-ray source
HESS~J1813--178.  We demonstrate that the non-thermal radio emission
is due to a young shell-type supernova remnant (SNR) G12.8--0.0, and
constrain its distance to be greater than 4~kpc. The X-ray emission is
primarily non-thermal and is consistent with either an SNR shell or
unidentified pulsar/pulsar wind nebula origin; pulsed emission is not
detected in archival {\em ASCA} data.  A simple synchrotron+inverse
Compton model for the broadband emission assuming that all of the
emission arises from the SNR shell implies maximum energies of
$(30-450)(B/10 {\rm\mu~G})^{-0.5}$ TeV.  Further observations are
needed to confirm that the broadband emission has a common origin and
better constrain the X-ray spectrum.

\end{abstract}

\keywords {acceleration of particles -- supernova remnants --
  radio:ISM -- Xrays:ISM -- ISM: individual (G12.8-0.0, HESS
  J1813-178)}

\section{INTRODUCTION}

Based on theoretical arguments, it has become widely accepted that a
significant fraction of Galactic cosmic rays with energies up to the
``knee'' in the cosmic ray spectrum at $\sim 10^{15}$~eV are generated
in shell type supernova remnants (SNRs)
\citep[e.g.][]{Blandford1987}. While the detection of non-thermal
X-rays from the shells of several SNRs provides evidence that SNR
shocks are efficient accelerators of $\la 10^{13}$ eV electrons
\citep[see e.g.][]{Reynolds1996}, direct evidence is still lacking
that (1) SNRs can accelerate particles all the way to $10^{15}$~eV,
and (2) SNRs efficiently accelerate protons and heavy nuclei as well
as electrons \citep[see][for a review]{Pohl2001}.  Powerful
constraints on shock acceleration in SNRs can come from the detection
of TeV \gr\/s.  However, these high energy photons can be
produced by a variety of mechanisms, including inverse Compton (IC)
emission from energetic electrons, and pion decay produced by the
collision of energetic protons with surrounding dense
gas. Additionally, pulsars, pulsar wind nebulae (PWNe), microquasars,
and massive stars, have also been suggested as TeV counterparts, so
it is unclear whether SNRs are in fact the primary source of
cosmic rays $\la 10^{15}$~eV.

Currently only two convincing shell SNR/TeV associations are known
\citep[SNRs~G347.3--0.5 and G266.2--1.2;][]{ah2004,ah2005a}, and
observations do not yet unambiguously distinguish between hadronic and
leptonic acceleration \citep{Lazendic2004,Malkov2005}.  However, with
the advent of new sensitive \v{C}erenkov imaging
telescopes, there are bright prospects for discovering additional TeV
sources.  Indeed, \citet{ah2005b} recently reported the discovery of
eight new TeV \gr\/ sources in the inner Galactic plane ($\ell=\pm
30\arcdeg$) using the High Energy Stereoscopic System (HESS).  Of the
eight new TeV sources, six are listed in \citet{ah2005b} as being in
close proximity to either an SNR, pulsar, or {\em EGRET} source (in
some cases all three).  However, these authors were unable to identify
any plausible counterparts to two of the new TeV sources and suggest
that they may represent a new class of ``dark'' nucleonic particle
accelerators.

In this {\em Letter} we present evidence that one of the two
``dark'' TeV sources, \hess\/, is positionally coincident with a
previously unidentified young radio and X-ray SNR, G12.8--0.0. 
A simple model for the broadband emission is also presented. 

\section{DATA AND RESULTS}

\subsection{The Radio Source \RS\/}

The radio source \RS\/ was discovered in a new low frequency Very
Large Array (VLA) 90cm survey incorporating B+C+D configuration data
\citep{Brogan2005}. The previously unidentified radio source has a
shell morphology with a diameter of $\sim 2.5\arcmin$. 
Analysis of archival VLA C+D configuration 20cm data has confirmed
this discovery, and indicates that the radio emission is
non-thermal. \RS\/ has integrated radio flux densities at 20 and 90 cm
of $0.65\pm 0.1$ and $1.2\pm 0.08$ Jy, respectively.  Figure 1a shows
the 90cm image while Figure 1b shows a 3-color image of the
region using 90cm, 20cm, and 8 \mum\/ {\em Spitzer Space Telescope}
GLIMPSE data \citep{Benjamin2003}. It is notable that no distinct
(non-diffuse) dust emission is coincident with \RS\/, confirming its
non-thermal nature. \hess\/ (positional uncertainty $1\arcmin-2\arcmin$) is
coincident with the radio shell (see Fig.\ 1a) and its size of $\sim
3\arcmin$ is in excellent agreement with that of \RS\/. No
structure is evident in the published $3\arcmin$ resolution HESS image
\citep{ah2005b}. The radio and TeV sources are located
$\sim8\arcmin$ from the \HII\/ region complex W33 (see Fig.\ 1a,b,
\S3.1).  

Though not previously recognized as a distinct source, \RS\/ is
evident as a faint unresolved extension to W33 in the Nobeyama 3cm,
Parkes 6cm, and Bonn 11cm surveys, with resolutions ranging from
$3\arcmin$ to $4.3\arcmin$ \citep{Haynes1978,Handa1987,Reich1990}.
The radio spectrum of \RS\/ (including the 20 and 90cm VLA data, 4m
VLA B+C+D configuration data from \citet{Brogan2004}, and the single
dish data) is shown in Figure 1c.  The rms noise in the 20 and
90cm images are 5 and 2.5 \mjb\/, respectively, and the flux density
uncertainty is ($\#$ independent beams)$^{0.5}\times 3\sigma$. \RS\/
is not detected at 4m; the flux density shown in Figure 1c is a
$5\sigma$ upper limit (also see \S3.1).  To account for the extra
uncertainty due to confusion with W33, $6\sigma$ was used for the
single dish uncertainties. The spectral index using these data
(excluding the 4m non-detection) and a weighted least squares fit is
$-0.48\pm 0.03$ (where $S_{\nu}\propto\nu^{\alpha}$), similar to that
of other small shell type SNRs (i.e.  G349.7+0.2 and W49B) without
central pulsars \citep{Green2004}.

Assuming $\alpha=-0.48$, the 1~GHz surface brightness of \RS\/ is
$\Sigma_{1{\rm GHz}}\sim 3.3\times 10^{-20}$ W~m$^{-2}$~Hz~sr$^{-1}$,
and the radio luminosity from $10^7-10^{11}$ Hz is $\sim 3.6
(d/d_4)^2\times 10^{32}$~\ergs\/ ($d$ is the distance and $d_4=4$~kpc;
see \S3.1).  The radio morphology, radio spectrum, and lack of
coincident mid-infrared emission (Fig.\ 1) provide convincing evidence
that \RS\/ is a previously unidentified shell type SNR; henceforth
designated \gt\/.

\subsection{The X-ray Source \AX\/}

Analysis of a pointed 100 ks archival {\em ASCA} observation toward
this region (taken 1997~September) reveals a source, hereafter \AX\/,
spatially coincident with both \gt\/ and \hess\/ (see Figure 2). \AX\/
is detected by both the SIS and GIS {\em ASCA} instruments, although
the emission is contaminated in the GIS detectors by the bright X-ray
binary GX~13+1 $\sim 40\arcmin$ to the NE. Thus, we have confined our
spectral and imaging analysis to the SIS data with a combined
(screened) exposure time of $\sim 64$ ks.  The X-ray source appears to
be either unresolved or slightly extended (Fig 2).  The X-ray peak
nominally lies interior to the radio shell, but since the {\em ASCA}
pointing uncertainty can be as large as $1\arcmin$
\citep{Gotthelf2000}, the {\em ASCA} image does not distinguish
between X-rays originating from the center (i.e. a compact object) or
the shell of the SNR.

The X-ray spectrum is quite hard, with emission extending to 10 keV
and a strong cut-off below 2 keV; no lines are evident (inset
Fig. 2). Thus it is clear that the emission is predominately
non-thermal. The background-corrected SIS data (extracted from a
$3\arcmin$ source region) consisting of $\approx 2000$ counts in the
energy range 0.5--10 keV were jointly fit to several absorbed spectral
models including a power-law, and 2-component combinations of a
power-law plus a (1) Raymond-Smith (RS) thermal plasma (with and
without the abundances frozen at solar), (2) blackbody, and (3)
Bremsstrahlung component.  The data quality are such that all of these
models give reasonable fits (reduced $\chi^2\sim 1$).

Statistically, a power-law plus RS model gives the best fit (RS
$kT\sim~0.22\pm0.1$~keV) but the improvement is only $4\sigma$ 
compared to a power-law alone; the emission measure of the RS component
cannot be constrained. This result suggests that higher quality X-ray
data are needed to determine the thermal contribution to the X-ray
emission \citep[see also][]{Ubertini2005}.  None of the two-component
fits yield a significantly different absorbing column or photon index
compared to the power law fit alone: $N_H =
(10.8^{-1.9}_{+2.3}\times 10^{22}$~cm$^{-2}$ and $\Gamma =
1.83^{-0.37}_{+0.42}$. The unabsorbed flux and luminosity of \AX\/
from 2--10~keV are $\sim 7.0\times10^{-12}$~ergs~cm$^{-2}$~s$^{-1}$
and $1.7 (d/d_4)^2\times 10^{34}$~\ergs\/, respectively. These
parameters are consistent with \AX\/ being either a pulsar/PWN or one
of a small number of primarily non-thermal X-ray SNRs such as
G266.2--1.2 and G347.3--0.5.

\section{DISCUSSION}

\subsection{Distance Constraints}

As mentioned previously, \gt\/ lies near the line of sight of the W33
\HII\/ region/star formation complex.  The radio
recombination lines, molecular lines (including masers), and \HI\/ gas
associated with W33 arise from LSR velocity components at $\sim +30$
and $\sim +50$~\kms\/
\citep[e.g.][]{Sato1977,Bieging1978,Gardner1983}. No \HI\/ absorption
features are evident toward W33 beyond $\sim +55$ \kms\/
\citep{Radhakrishnan1972}. Since the tangent point velocity is $\sim
+170$ \kms\/ in this direction \citep[assumes Galactic center distance
8.5 kpc and][rotation curve]{Fich1989}, W33 must lie at about the
$+55$ \kms\/ near distance of $\sim 4.3$ kpc. Similar to W33, \HI\/
absorption spectra from the Southern Galactic Plane Survey
\citep[resolution $1\arcmin$, see][]{MG2005} toward G12.8--0.0
show absorption to, but not beyond $\sim +55$ \kms\/, although the
signal-to-noise is low. Thus, the distance to G12.8--0.0 is likely to
be $\ga 4$~kpc.

Based on the derived radio spectral index ($-0.48$) the expected 4m
flux density is $\sim 2.7$ Jy, suggesting that \gt\/ should be readily
detectable, yet it is not (Fig.\ 1c, \S2.1).  This discrepancy could
be due to the close proximity of W33 if G12.8--0.0 lies immersed in or
behind the \HII\/ region due to free-free absorption (see e.g Kassim
1989). In interferometric low frequency images ($\nu\la 100$ MHz)
\HII\/ regions can appear in absorption against the resolved out
diffuse Galactic plane synchrotron emission due to free-free
absorption.  W33 is coincident with such a 4m
absorption 'hole'; its morphology is similar to the diffuse
90cm and 8 \mum\/ emission (Fig.\ 1b) and extends as far
west as G12.8--0.0.  Thus, free-free absorption is the most natural
explanation for the non-detection of G12.8--0.0 at 4m, and confirms
the \HI\/ result that the SNR must lie at the distance of or behind
W33.

The high column density derived from the {\em ASCA}\ data, $N_H
\sim 10^{23}$~\ct\/, also suggests that \gt\/ lies behind W33. Along
nearby lines of sight, the integrated Galactic \HI\ column density is
only $\sim 2\times 10^{22}$~\ct\/ \citep{Dickey1990}, suggesting that
a significant discrete source of absorption must be in the foreground
to the SNR.  CO data \citep{Dame2001} indicate that dense gas in the
distance range 0--4~kpc (much of which is associated with W33)
contributes a hydrogen column density $\sim 8\times 10^{22}$~\ct\/,
and can thus account for much of the X-ray absorption. This further
argues for a distance of 4~kpc or greater for \gt\/; we adopt a
fiducial minimum distance of 4~kpc ($d_4$) for the remaining
discussion.

\subsection{The Probable Youth of \gt\/}

The small angular size of \gt\/ ($2.5\arcmin$) suggests that it is
quite young; certainly the high extinction toward this region would
have prevented its detection optically. The radius of \gt\/ is
$1.5(d/d_4)$ pc.  If the SNR is still freely expanding at a speed
$v_s$ its age is $\sim 285(d/d_4)~(v_s/v_5)^{-1}$ years ($v_5=5,000$
\kms\/). Shock speeds of $v_s=5,000$ \kms\/ are expected from SNRs
with X-ray synchrotron emission \citep[e.g.][]{aa1999}. If the
distance is 4--10~kpc, use of the free-expansion age is justified
since the swept up mass would still be a small fraction of the ejected
mass.  At greater distances the SNR is likely to have entered the
Sedov-Taylor phase. For example, at 20 kpc, the swept up mass is $\sim
50$ M$_{\sun}$ (assuming an ambient density $n_o= 1$ \cc\/ and $10\%$
helium abundance) and the Sedov-Taylor age is
$2,520(n_o)^{0.5}(E/E_{51})^{-0.5}$ years (where $E$ is the kinetic
energy and $E_{51}=10^{51}$ ergs).  Thus, independent of distance, if
G12.8--0.0 is expanding into a typical density medium it is very young
with representative age estimates in the range 285-2,500
years. Excluding G12.8--0.0, only six shell type Galactic SNRs with
diameters $\la 3\arcmin$ are currently known, undoubtedly due to
observational selection effects since many more are expected
\citep{Green2004}.  Our discovery highlights the effectiveness of high
resolution and sensitivity low radio frequency surveys in finding
young Galactic SNRs.

\subsection{X-ray and TeV Emission from a Young Pulsar?}

Although there is excellent positional agreement between the radio,
X-ray, and TeV emission (Fig.\ 2), it is currently unclear whether all
these components originate from the SNR shell or if there is an as yet
unidentified associated pulsar or PWN.  No central radio nebula or
point source is detected at our current spatial resolution and
sensitivity, and there are no known radio pulsars within $16\arcmin$
of the SNR \citep{Manchester2005}. This field has been searched for
pulsars as part of the Parkes multi-beam pulsar survey, to a
limiting 1.4~GHz flux density of 0.2 mJy \citep{Manchester2001},
corresponding to a pseudo-luminosity of $\sim3(d/d_4)^2$~mJy~kpc$^2$.
Recent deep pulsar searches toward SNRs have been able to either
detect or place upper limits of $\sim 1$~mJy~kpc$^2$ on the 1.4 GHz
pseudo-luminosity of central pulsars \citep[e.g.][]{Camilo2002}.
Thus, the Parkes multi-beam survey is reasonably constraining, but a
deeper search is needed.

The X-ray luminosity and photon index inferred in \S2.2 for \AX\/ are
typical of young pulsars and their associated PWN
\citep[e.g.][]{Possenti2002}. In order to search for X-ray pulsations
from \AX\/ we conducted a $Z_n^2$ search \citep[see][]{Buccheri1983}
on the barycenter-corrected {\em ASCA} high-bit mode GIS data (with 20
ks of data).  No pulsed signal was detected using 585 counts between
4--8 keV.  For periods between 125~ms and 1000~s, we find an upper
limit on the pulsed fraction of $38\%$ ($19\%$) for a sinusoidal
(sharply peaked) pulse profile.  Since most young pulsars rotate with
periods below this range, and the pulsed signal may be a small
fraction ($<10\%$) of the total pulsar plus PWN X-ray flux, the lack
of pulsations is not very constraining. Deeper X-ray and radio
observations of this source are needed to confirm or rule out a
pulsar/PWN interpretation for \AX\/.

\subsection{X-ray and TeV Emission from the SNR Shell?}

The radio (SNR shell) and X-ray (origin uncertain) data fit smoothly
onto a single power-law+rolloff spectrum, unlike the typical behavior
of PWN, suggesting they have a common origin (Figure 3).  Assuming
that the radio, X-ray, and TeV emission originate from the SNR shell,
we have modeled the flux expected from synchrotron+IC mechanisms using
a power law energy distribution modified by an exponential cutoff
$dN/dE\approx~E^{-\sigma} {\rm exp}[-(E/E_{max})]$ \citep[where
$\sigma$ is the index of the electron distribution, and $E_{max}$ is
the maximum energy of accelerated particles, see][]{Lazendic2004}.
Since the X-ray absorption column (and hence X-ray spectral index) is
not well constrained by the {\em ASCA} data, two variations of this
model are shown in Figure 3, one using the best fit X-ray $N_{\rm H}$
(red) and the other using the $1\sigma$ lower limit to $N_{\rm H}$
(blue). Together these models encompass the likely range of parameter
space. We have assumed a $15\%$ filling factor for the magnetic field
in the IC emitting region in order to obtain the best agreement with
the HESS data for the `blue' model; larger values yield lower IC
components. The fitted broad-band electron spectral indices are
$\sigma=2.3$ (red) and $\sigma=2.0$ (blue) -- the former is consistent
with the best-fit photon index for the {\em ASCA} spectrum,
while the latter is more consistent with the radio spectral index.

The roll-off frequencies ($\nu_{ro}$) derived from the synchrotron
spectra are $3.3\times10^{19}$~Hz (red) and $1.3\times10^{17}$~Hz
(blue).  The maximum energy is related to $\nu_{ro}$ and the magnetic
field strength $B$ by $\nu_{ro}\sim 1.6\times
10^{16}~(B/10~{\rm\mu~G})~(E_{max}/10~{\rm TeV})^2$ Hz, so that
$E_{max}=450~(B/10 {\rm\mu~G})^{-0.5}$ TeV (red) and $E_{max}=30~(B/10
{\rm\mu~G})^{-0.5}$ TeV (blue). Thus, unless the magnetic field
strength is much stronger than 10 \muG\/ \citep[see for
example][]{Volk2005}, $E_{max}$ for this source is potentially rather
high \citep[e.g.][]{Reynolds1999}. Additionally, for much of the
parameter space range, the IC emission from CMB scattering alone is
insufficient to account for the HESS data; the discrepancy is larger
for larger values of $B$.  It is possible a more sophisticated IC
model incorporating scattering off starlight \citep{ah2005c} would
better accommodate the HESS data. Indeed, \citet{Helfand2005} find
this to be the case assuming the SNR lies at the distance
of W33, though currently it is only constrained to be at or
greater than this distance.  It is also possible that the TeV emission
arises from the acceleration of nuclei and subsequent pion decay
instead of IC if the SNR is embedded in a sufficiently dense ambient
medium \citep[e.g.][]{Uchiyama2005}.  Detection of resolved X-ray and
TeV emission from the SNR shell would strengthen our suggestions that
the radio, X-ray, and TeV emission have the same origin. Higher
sensitivity X-ray data will also be crucial to better constrain the
properties of the broadband emission.

\acknowledgments 

Basic research in radio astronomy at the Naval Research Laboratory is
supported by the Office of Naval Research. The VLA is operated by the
National Radio Astronomy Observatory -- a facility of the National
Science Foundation operated under cooperative agreement by Associated
Universities, Inc.  This research made use of HEASARC, provided by NASA's
Goddard Space Flight Center. We thank L.~Rickard, D. Ellison, and
S. Reynolds for helpful discussions.


\begin{figure}[h!] 
\epsscale{1}
\plotone{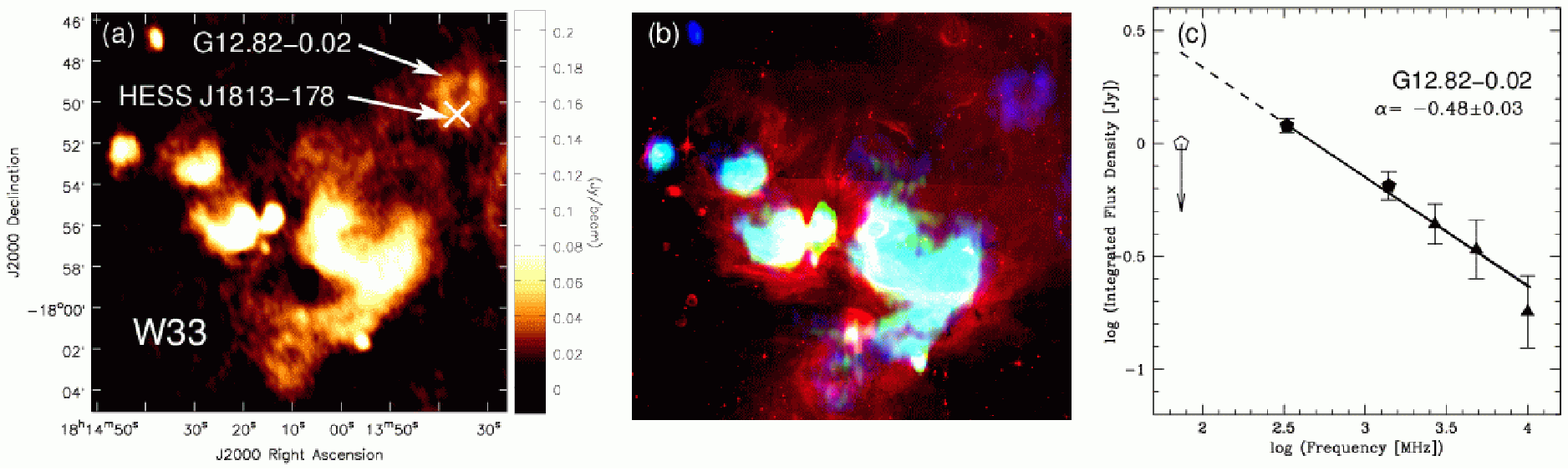}
\caption{(a) VLA 90cm image of \RS\/ with $32\arcsec\times 19\arcsec$
resolution. The cross indicates the position of \hess\/ which has a
positional uncertainty of $1\arcmin-2\arcmin$ \citep{ah2005b}. (b)
Three color image of the same region as (a), with red={\em Spitzer} 8
\mum\/, green=VLA 20cm, and blue=VLA 90cm. Thermal emission is white
to cyan or red in color, while non-thermal emission is darker
blue. (c) Radio spectrum for \gt\/. Solid hexagon symbols are from the
current work at 20 and 90cm. The open hexagon symbol indicates the
$5\sigma$ 4m upper limit. The $\blacktriangle$ symbols indicate flux
densities from the 11cm Bonn, 6cm Parkes, and 3cm Nobeyama single dish
surveys \citep{Reich1990,Haynes1978,Handa1987}.  The line (solid and
dashed) shows a power-law weighted least squares fit to the radio data
for $\nu\ge 330$ MHz (90cm).}
\end{figure} 


\begin{figure}[h!] 
\epsscale{0.5}
\plotone{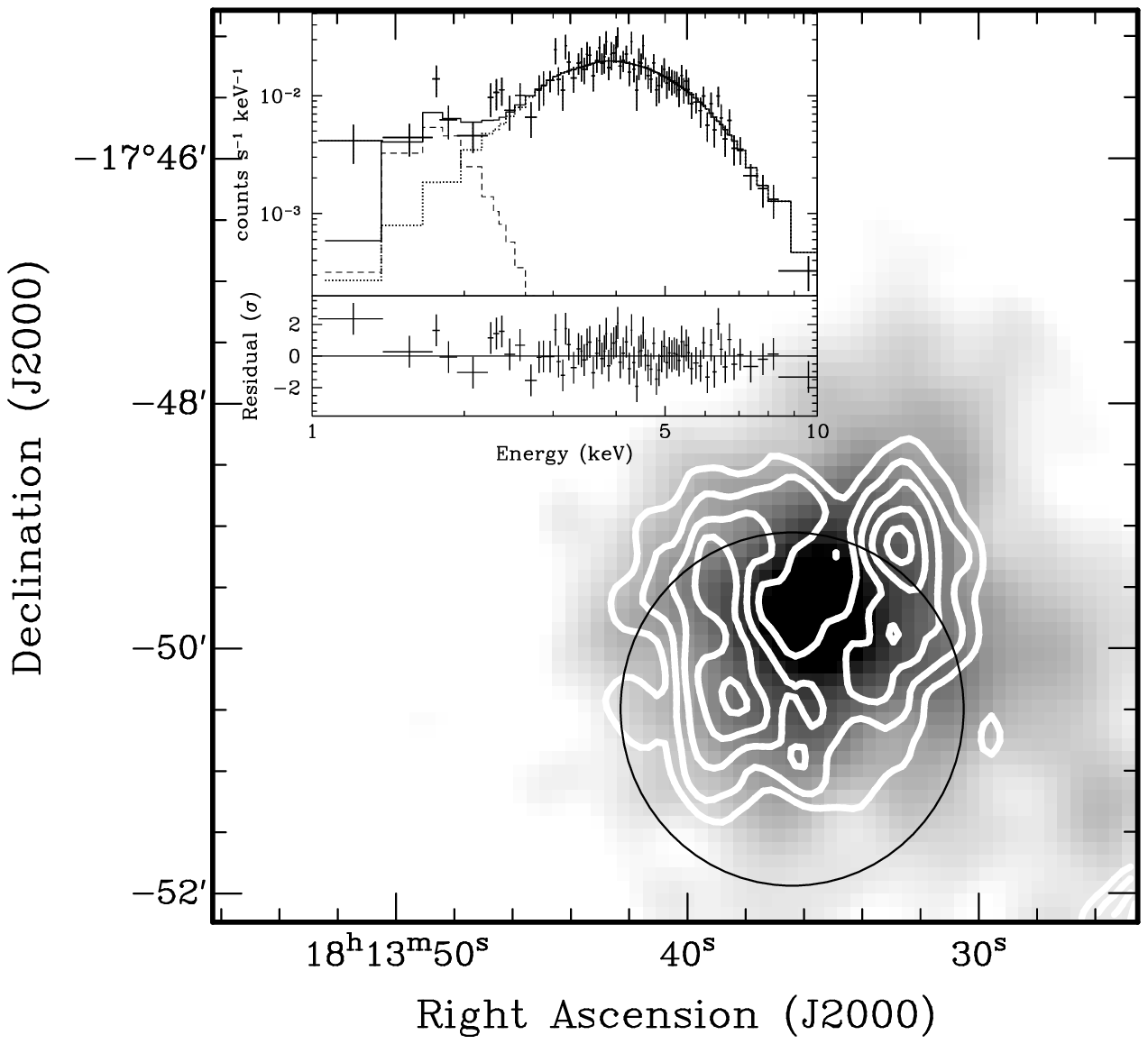}

\caption{{\em ASCA} SIS greyscale image of \AX\/ in the energy range
  0.5--10 keV smoothed to $30\arcsec$ with 90cm contours from Fig. 1a
  at 28, 33, 38, 43, and 48 \mjb\/ overlaid (contour intervals are
  $2\sigma$).  The black circle indicates the position and angular
  extent ($3\arcmin$) of \hess\/. {\em Inset} {\em ASCA} SIS spectrum
  of \AX\/ extracted from $3\arcmin$ region encompassing the radio
  shell. The cross symbols show the data while the solid line shows
  the best fit power-law (dotted line) plus RS (dashed) model. The
  residuals in the form of the uncertainty $\sigma$ are shown in the
  bottom panel.}

\end{figure} 


\begin{figure}[h!] 
\epsscale{0.5}
\plotone{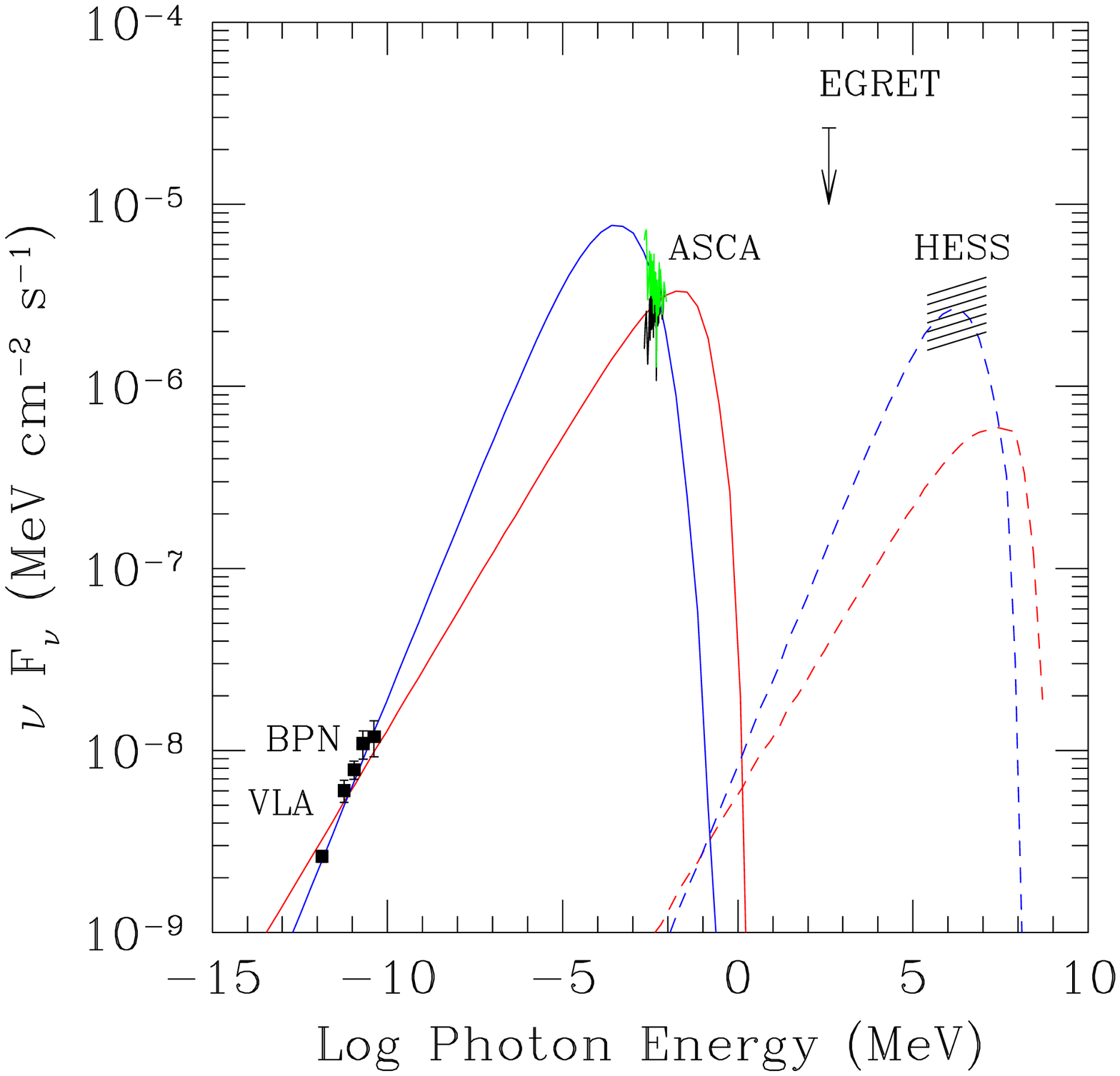}
\caption{Fits to the broadband emission assuming that all the flux
originates from the shell of \gt\/. BPN stands for Bonn, Parkes, and
Nobeyama. The diagonal black lines indicate both the uncertainty in
the HESS flux measurements, and the fact that no spectral information
has yet been published for the TeV emission. The two models indicated
by the red and blue lines show the range of parameter space that best
fit the data: the {\em red} model uses the spectral index from the
best fit to the {\em ASCA} data $N_{\rm H}$ of $10.8\times 10^{22}$ cm$^{-2}$
(black X-ray spectrum), while the {\em blue} model uses the spectral
index implied by the $1\sigma$ lower limit to $N_{\rm H}$ of
$8.9\times 10^{22}$ cm$^{-2}$ (green X-ray spectrum). Both models
include contributions from synchrotron (solid lines) and IC (dashed
lines) mechanisms; we have assumed that the filling factor of the
magnetic field in the IC emitting region is $15\%$.  }
\end{figure} 


\begin{thebibliography}

\bibitem[Aharonian et al.(2005a)]{ah2005a} Aharonian, F., et 
al.\ 2005a, \aap, in press, astro-ph/0505380

\bibitem[Aharonian et al.(2005b)]{ah2005b} Aharonian, F., et 
al.\ 2005b, Science, 307, 1938

\bibitem[Aharonian et al.(2005c)]{ah2005c} Aharonian, F., et 
al.\ 2005c, \aap, 432, L25

\bibitem[Aharonian et al.(2004)]{ah2004} Aharonian, F.~A., et 
al.\ 2004, \nat, 432, 75  

\bibitem[Aharonian \& Atoyan(1999)]{aa1999} Aharonian, F.~A., 
\& Atoyan, A.~M.\ 1999, \aap, 351, 330 

\bibitem[Benjamin et al.(2003)]{Benjamin2003} Benjamin, R.~A., et 
al.\ 2003, \pasp, 115, 953 

\bibitem[Bieging et al.(1978)]{Bieging1978} Bieging, J.~H., 
Pankonin, V., \& Smith, L.~F.\ 1978, \aap, 64, 341 

\bibitem[Blandford \& Eichler(1987)]{Blandford1987} Blandford, R., \& 
Eichler, D.\ 1987, \physrep, 154, 1 

\bibitem[Brogan et al.(2004)]{Brogan2004} Brogan, C.~L., et al.\ 2004, 
\aj, 127, 355 

\bibitem[Brogan et al.(2005), in prep.]{Brogan2005} Brogan, C.~L., Gelfand,
J.~D., Gaensler, B.~M., Kassim, N.~E., \& Lazio, T.~J., 2005, in prep.

\bibitem[Buccheri et al.(1983)]{Buccheri1983} Buccheri, R., et al.\ 
1983, \aap, 128, 245 

\bibitem[Camilo et al.(2002)]{Camilo2002} Camilo, F., Manchester, 
R.~N., Gaensler, B.~M., \& Lorimer, D.~R.\ 2002, \apjl, 579, L25 


\bibitem[Dame et al.(2001)]{Dame2001} Dame, T.~M., Hartmann, D., 
\& Thaddeus, P.\ 2001, \apj, 547, 792 

\bibitem[Dickey \& Lockman(1990)]{Dickey1990} Dickey, J.~M., \& 
Lockman, F.~J.\ 1990, \araa, 28, 215 


\bibitem[Fich et al.(1989)]{Fich1989} Fich, M., Blitz, L., \& 
Stark, A.~A.\ 1989, \apj, 342, 272

\bibitem[Gardner et al.(1983)]{Gardner1983} Gardner, F.~F., 
Whiteoak, J.~B., \& Otrupcek, R.~E.\ 1983, Proceedings of the Astronomical 
Society of Australia, 5, 221 

\bibitem[Green (2004)]{Green2004} Green, D.~A. \ 2004, Bull. Astro. Soc. 
India, 32, 335

\bibitem[Gotthelf et al.(2000)]{Gotthelf2000} Gotthelf, E.~V., Ueda, 
Y., Fujimoto, R., Kii, T., \& Yamaoka, K.\ 2000, \apj, 543, 417 

\bibitem[Helfand et al.(2005)]{Helfand2005} Helfand, D.~J., Becker, 
R.~H., \& White, R.~L.\ 2005, astro-ph/0505392 

  \bibitem[Handa et al.(1987)]{Handa1987} Handa, T., Sofue, Y., 
Nakai, N., Hirabayashi, H., \& Inoue, M.\ 1987, \pasj, 39, 709 

\bibitem[Haynes et al.(1978)]{Haynes1978} Haynes, R.~F., Caswell, 
J.~L., \& Simons, L.~W.~J.\ 1978, Australian Journal of Physics 
Astrophysical Supplement, 45, 1 


\bibitem[Kassim(1989)]{Kassim1989} Kassim, N.~E.\ 1989, \apj, 347,
915
\bibitem[Lazendic et al.(2004)]{Lazendic2004} Lazendic, J.~S., 
Slane, P.~O., Gaensler, B.~M., Reynolds, S.~P., Plucinsky, P.~P., \& 
Hughes, J.~P.\ 2004, \apj, 602, 271 

\bibitem[Malkov, Diamond, \& Sagdeev (2005)]{Malkov2005} Malkov,
M.~A., Diamond, P.~H., \& Sagdeev, R.~Z. \ 2005, \apj, 624, L37

\bibitem[Manchester et al.(2001)]{Manchester2001} Manchester, R.~N., 
et al.\ 2001, \mnras, 328, 17 

\bibitem[Manchester et al.(2005)]{Manchester2005} Manchester, R.~N., 
Hobbs, G.~B., Teoh, A., \& Hobbs, M.\ 2005, \aj, 129, 1993 

\bibitem[McClure-Griffiths et al.(2005)]{MG2005} McClure-Griffiths,
et. al.\ 2005, ApJS, in press, astro-ph/0503134

\bibitem[Pohl (2001)]{Pohl2001} Pohl, M.\ 2001, {\em ``27th International 
Cosmic Ray Conference''} ed. Schlickeiser, R., Copernicus Group,
astro-ph/0111552

\bibitem[Possenti et al.(2002)]{Possenti2002} Possenti, A., Cerutti, 
R., Colpi, M., \& Mereghetti, S.\ 2002, \aap, 387, 993 

\bibitem[Radhakrishnan et al.(1972)]{Radhakrishnan1972} Radhakrishnan, 
V., Goss, W.~M., Murray, J.~D., \& Brooks, J.~W.\ 1972, \apjs, 24, 49 

\bibitem[Reich et al.(1990)]{Reich1990} Reich, W., Fuerst, E., 
Reich, P., \& Reif, K.\ 1990, \aaps, 85, 633 

\bibitem[Reynolds(1996)]{Reynolds1996} Reynolds, S.~P.\ 1996, \apjl, 
459, L13 

\bibitem[Reynolds \& Keohane(1999)]{Reynolds1999} Reynolds, S.~P., 
\& Keohane, J.~W.\ 1999, \apj, 525, 368 
 
\bibitem[Sato(1977)]{Sato1977} Sato, F.\ 1977, \pasj, 29, 75 
 
\bibitem[V{\"o}lk et al.(2005)]{Volk2005} V{\"o}lk, H.~J., 
Berezhko, E.~G., \& Ksenofontov, L.~T.\ 2005, \aap, 433, 229

\bibitem[Ubertini et al.(2005)]{Ubertini2005} Ubertini et al.\ 2005, 
astro-ph/0505191

\bibitem[Uchiyama et al.(2005)]{Uchiyama2005} Uchiyama, Y., 
Aharonian, F.~A., Takahashi, T., Hiraga, J.~S., Moriguchi, Y., \& Fukui, 
Y.\ 2005, AIP Conf.~Proc.~745: High Energy Gamma-Ray Astronomy, 745, 305 
 
\end{thebibliography}
\end{document}